\documentclass[aps,twocolumn,showpacs]{revtex4}
\usepackage{graphicx}
\usepackage{amsmath}
\usepackage{amsfonts}
\usepackage{color}
\newcommand{\bra}{\left\langle}
\newcommand{\ket}{\right\rangle}
\providecommand{\abs}[1]{\lvert#1\rvert}

\begin{document}

\title{Topological and spectral properties of random digraphs} 

\author{C. T. Mart\'inez-Mart\'inez,$^{1}$ J. A. M\'endez-Berm\'udez,$^2$ and Jos\'e M. Sigarreta$^{1}$}

\address{$^1$Universidad Aut\'onoma de Guerrero, Centro Acapulco CP 39610, Acapulco de Ju\'arez, Guerrero, Mexico\\$^2$Instituto de F\'{\i}sica, Benem\'erita Universidad Aut\'onoma de Puebla, Puebla 72570, Mexico 
}

\date{\today} \widetext

%\pacs{05.45.-a, 05.45.Pq, 05.45.Tp}

\begin{abstract}

We investigate some topological and spectral properties of Erd\H{o}s-R\'{e}nyi (ER) random digraphs $D(n,p)$. 
In terms of topological properties, our primary focus lies in analyzing the number of non-isolated vertices 
$V_x(D)$ as well as two vertex-degree-based topological indices: the Randi\'c index $R(D)$ and sum-connectivity 
index $\chi(D)$. 
First, by performing a scaling analysis we show that the average degree $\bra k \ket$ serves as scaling 
parameter for the average values of $V_x(D)$, $R(D)$ and $\chi(D)$.
Then, we also state expressions relating the number of arcs, spectral radius, and closed walks of length 2 to 
$(n,p)$, the parameters of ER random digraphs. 
Concerning spectral properties, we compute six different graph energies on $D(n,p)$. 
We start by validating $\bra k \ket$ as the scaling parameter of the graph energies. 
Additionally, we reformulate a set of bounds previously reported in the literature for these energies as a function 
$(n,p)$. Finally, we phenomenologically state relations between energies that allow us to extend previously 
known bounds.

\end{abstract}

\maketitle

\section{Introduction}

In recent years, there has been a significant increase in the use of graphs to represent complex systems in various fields, including computer science, engineering, biology, and social sciences~\cite{B16, N10, BA99, AB02, GMDMSP18}. This growing trend can be attributed to the effectiveness of capturing the properties of complex systems through graphs,  where the vertices represent the agents of the system and the edges reflect their interactions. This, in turn, opens the door to the analysis of complex systems through various mathematical techniques coming mainly from graph theory. 

The study of the properties of graphs covers many aspects, focusing mainly on topological and spectral properties. One of the ways to study and characterize these properties is through their topological descriptors, such as degree distribution, clustering coefficient, eigenvector centrality, energy, and, more recently, topological indices~\cite{N03, LM01, NMHPLB18, MZN14}.

Although many studies have been carried out with highly relevant results about the topological and spectral properties of graphs, most of them focus on graphs whose edges do not have a specific direction, in which the connection between two vertices is symmetric and bidirectional (undirected graphs). However, in several cases, it is mandatory to incorporate the direction of the information flow when considering the modeling of real-world systems. This is indeed the case when considering food webs~\cite{ABB05, L80, ZU03}, neural networks~\cite{SZ04, SCKH04, RS10}, genetic regulation~\cite{BGL11}, chemical networks~\cite{VRJ13, PDSC12}, fluid flows~\cite{SMCN10, K021} or financial networks~\cite{AOT15}, among many other relevant applications. In these scenarios, it is crucial to capture the orientation of the connections so that the systems are represented by directed networks, commonly known as digraphs. Consequently, there is a specific interest in exploring the properties of directed graphs.

A digraph or directed graph is a mathematical structure denoted as $D=(V, E)$, where $V$ represents a finite set of $n$ elements called vertices or nodes and $E \subset V \times V$ comprises $m$ directed edges (also called arcs) connecting vertices. 

Topological properties delve into the fundamental structural properties of digraphs, including connectivity, accessibility, cycles, and paths. In this line, applying topological indices based on vertex degrees to characterize and analyze the topological properties of graphs has been a widely used approach. The concept of an index based on vertex degrees originates in chemical graph theory, which uses graph theory to study the properties of chemical compounds by representing them as graphs, where atoms are vertices and bonds are edges. The vertex-degree-based (VDB)  topological indices quantify some aspects of the topology of the graph in relation to the degrees of its vertices.
In a general formulation, a VDB topological index can be expressed as~\cite{G13}:
\begin{equation}
\label{gen-inx}
TI=TI(G)=\sum_{i \sim j} f(k_i,k_j),
\end{equation}
where the summation extends over all pairs of adjacent vertices, denoted as $i$ and $j$, within the molecular graph $G$, $k_{i}$ is the degree of the vertex $i$ and $f(k_i,k_j)$ represents a function tailored to the specific topological property under investigation. Since applying these indices to the study and characterization of the topological properties of graphs has acquired great relevance, many topological indices have been proposed. However, extending this concept to directed graphs is a complex task since, in digraphs, each vertex has an out-degree, an in-degree, and a total degree. However, Monsalve and Rada have recently presented a generalization of VDB topological indices applied to digraphs~\cite{MR21}. Consequently, there are still few works in which the properties of these topological indices have been explored~\cite{MR21b, CMR22, AA22}.

On the other hand, the study of spectral properties involves the study of eigenvalues and eigenvectors associated with matrices corresponding to digraphs, such as the adjacency matrix, the Laplacian matrix, and the Hermitian matrix, among others. In this context, also rooted in chemical graph theory, the concept of energy emerges as a spectral quantity that serves as a descriptor of the properties of a graph and allows the characterization and study of the properties of specific systems. The concept of energy was initially introduced using the eigenvalues of the adjacency matrix associated with a graph: For a simple undirected graph, the adjacency matrix is defined through the matrix elements
\begin{equation}
A_{i j}=\left\{\begin{array}{ll}
1 & \text { if there is an edge between vertices $i$ and $j$, } \\
0 & \text { otherwise. }
\end{array}\right.
\end{equation}
 In 1978 Ivan Gutman proposed the concept of energy of a finite and undirected simple graph based on Huckel's orbital model as~\cite{G78, G01}
\begin{equation}
E(G)=\sum_{i=1}^{n} |\lambda_{i}|,
\label{Eq1}
\end{equation} 
where $\lambda_{i}$ are the eigenvalues of the adjacency matrix of the graph.
Furthermore, other energies associated with other graph matrices have been proposed, such as the Laplacian energy~\cite{GZ06}, the Laplacian-energy like invariant~\cite{LL08}, the signless Laplacian energy~\cite{SRAG}, the distance energy~\cite{IGV10}, the incidence energy~\cite{JKM09}, the skew energy~\cite{ABS10}, the Sombor energy~\cite{GN21}, the Randi\'c energy~\cite{BGGC10}, the Seidel energy~\cite{H12}, etc. 

Moreover, the Coulson integral is a complex integral that allows computing the energy of a graph without directly calculating its eigenvalues: Let $\phi$ be the characteristic polynomial of the adjacency matrix of the graph, then $\phi$ is the characteristic polynomial of the graph, which is defined as
\begin{equation}
\phi (G,x)=\det [xI-A(G)],
\end{equation}
where $I$ is the identity matrix of order $n$.
The Coulson integral is defined as~\cite{C40}
\begin{equation}
E(G)=\frac{1}{\pi}\int_{-\infty}^{\infty} \left( n-\frac{ix\phi ' (G, ix)}{\phi(G,ix)} \right) dx,
\end{equation}
where $\phi ' (G, ix)$ is the derivative of $\phi(G,x)$ and $n$ is the order of the adjacency matrix.
 
The energy of a graph has several applications in various fields, such as chemistry, physics, mathematics, biology, social networks,  computer science, etc~\cite{DME13, DGF12, XXZCG20, SSN17, KM19}. It is mainly used as an indicator of the graph structure that determines specific properties of the system represented by the graph or to optimize specific processes. Furthermore, the graph energy has been used as a criterion for graph classification. Depending on the value of their energy, graphs can be categorized as hyperenergetic if $E(G) > 2(n-1)$ or non-hyperenergetic if $E(G) \leq 2(n-1)$; the energy value of a complete graph serves as a reference in this sense~\cite{B04}. 

Thus, the interest in the study of the energy of a graph has grown significantly. The most notable results in this field focus mainly on determining upper and lower bounds for this magnitude based on various properties of the graphs, mainly of a topological nature. One of the most important bounds on the energy of a graph is the McClelland inequality, which establishes a relationship between the energy and the number of vertices and edges of the corresponding graph~\cite{M71}:
\begin{equation}
E(G) \leq \sqrt{2mn}.
\end{equation}

It is important to notice that for digraphs, the adjacency matrix is not necessarily symmetric, so its eigenvalues can be complex, and the definition of the graph energy of Eq.~(\ref{Eq1}) cannot be straightforwardly extended. Given this, several definitions of digraph energies have also been proposed and studied. 
Therefore, this work investigates topological and spectral characteristics of directed random graphs, focusing on the Erd\H{o}s-R\'{e}nyi model.

\section{Topological and spectral properties of Erd\H{o}s-R\'{e}nyi digraphs}
\label{topological} 

An Erd\H{o}s-R\'{e}nyi (ER) digraph, denoted by $D(n,p)$, is a directed random graph with $n$ independent vertices connected with probability $p$. Given two vertices $u$ and $v$, $p$ is the probability that there is an arc from vertex $u$ to vertex $v$, so $p \in (0,1)$. When $p=0$, the graph consists of $n$ isolated vertices; when $p=1$, it becomes a complete graph. We can obtain graphs between these two extremes by varying the value of $p$ between $0$ and $1$. It is important to note that for $0<p<1$, a given pair of parameters $(n,p)$ represents an infinity set of random graphs. Therefore, calculating a property for a single graph is not informative. Instead, we can obtain more relevant information by calculating a given average property over an ensemble of random graphs characterized by the same pair of parameters $(n,p)$. Although this statistical approach is a common practice in random matrix theory (RMT), it is not as common in graph theory; however, it has been applied recently to several random graph models~\cite{MMRS20, AHMS20, AMRS20, MMRS21, MAMRP15, PRRCM20,PM23}.

Thus, below, we perform a numerical analysis of some topological properties of ER digraphs by the use of the number of non-isolated vertices ($V_x(D)$) and the Randi\'c  ($R(D)$) and the sum-connectivity ($\chi(D)$) indices. 

Following the generalization of the concept of VDB topological indices of digraphs proposed by Monsalve and Rada~\cite{MR21}, the Randi\'c and the sum-connectivity indices are respectively defined as:
\begin{equation}
R(D)=\frac{1}{2}\sum_{uv \in D} \frac{1}{\sqrt{{k_{u}}^{+}{k_{v}}^{-}}}
\end{equation}
and
\begin{equation}
\chi(D)=\frac{1}{2}\sum_{uv \in D} \frac{1}{\sqrt{{k_{u}}^{+}+{k_{v}}^{-}}},
\end{equation}
where $uv$ denotes the arc connecting vertices $u$ and $v$, ${k_{u}}^{+}$ denotes the out-degree of
the vertex $u$, and ${k_{v}}^{-}$ denotes the in-degree of the vertex $v$.

\begin{figure}
\centering
\includegraphics[scale=0.4]{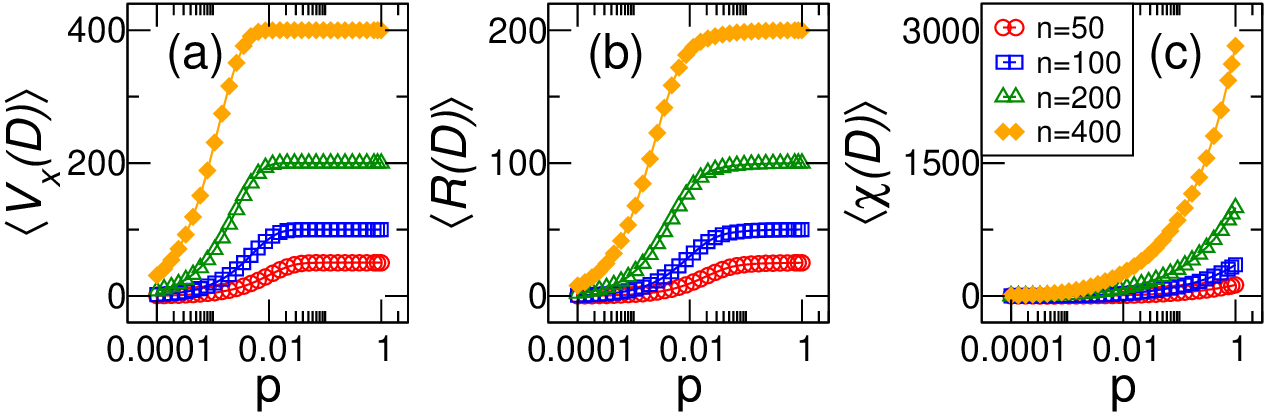}
\caption{(a) Average number of non-isolated vertices $\bra V_{x}(D) \ket$, (b) average Randi\'c index $\bra R(D) \ket$ and (c) average sum-connectivity index $\bra \chi(D) \ket$ as a function of the connection probability $p$ of Erd\H{o}s-R\'{e}nyi digraphs of different sizes $n \in [50,400]$. Each symbol was calculated by averaging over $10^6/n$ random digraphs $D(n,p)$.}
\label{Fig1}
\end{figure}

\begin{figure}
\centering
\includegraphics[scale=0.4]{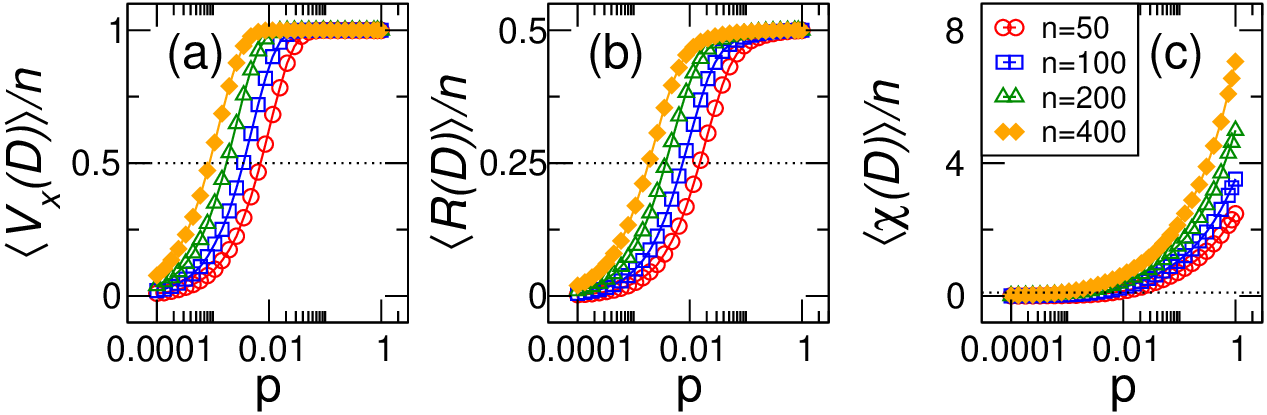}
\caption{(a) $\bra V_{x}(D) \ket$, (b) $\bra R(D) \ket$ and (c) $\bra \chi(D) \ket$ normalized to $n$ as a function of the connection probability $p$ of Erd\H{o}s-R\'{e}nyi digraphs of different sizes $n \in [50,400]$. Dotted lines in panels (a-c) correspond to $0.5$, $0.25$ and $0.1$, respectively. Same data sets of Fig.~\ref{Fig1}. }
\label{Fig2}
\end{figure}

First, we compute the average values of $V_x(D)$, $R(D)$ and $\chi(D)$ for ensembles of adjacency matrices of ER digraphs characterized by different combinations of parameters $(n,p)$. In Fig.~\ref{Fig1}, these quantities are shown for four different graph sizes as a function of the connection probability $p$. We can observe that the curves corresponding to each quantity follow similar shapes but are displaced in the $p$-axis depending on the graph size. To better appreciate the shape of these curves, we normalize them to the size of the network and plot them again in Fig.~\ref{Fig2}.

In Figs.~\ref{Fig2}(a) and~(b), it can be seen that the shape of the normalized curves of $\bra V_{x}(D) \ket$ and $\bra R(D) \ket$ are very similar. Initially, for small values of $p$, both are close to zero, then they increase with $p$ until they reach their maximum values. In the case of $\bra V_x(D) \ket/n$, the maximum value is $1$, while for $\bra R(D)\ket/n$ it is $1/2$. However, Fig.~\ref{Fig2}(c) shows a different picture for the normalized curves of $\bra \chi(D) \ket$. That is, $\bra \chi(D) \ket/n$ is a strictly monotone increasing function and its maximum value depends on the graph size. The maximum value of $\bra \chi(D) \ket/n$ is reached at $p=1$ and is equal to $\sqrt{(n-1)/8}$.

Notably, in all three cases, the curves corresponding to the same quantity exhibit a similar behavior but they are shifted along the $p$-axis for different graph sizes $n$. Now, our goal is to identify a scaling parameter for these quantities. To achieve this, we first need to quantify the displacement of the curves with $n$. Then, without loss of generality, we characterize the displacement by computing the value of $p$ (that we label as $p^*$) for which $\bra V_x(D) \ket/n$, $\bra R(D)\ket/n$ and $\bra \chi(D) \ket/n$ reach the value of 0.5, 0.25 and 0.1, respectively; see the dotted lines in Fig.~\ref{Fig2}.

\begin{figure}
\centering
\includegraphics[scale=0.4]{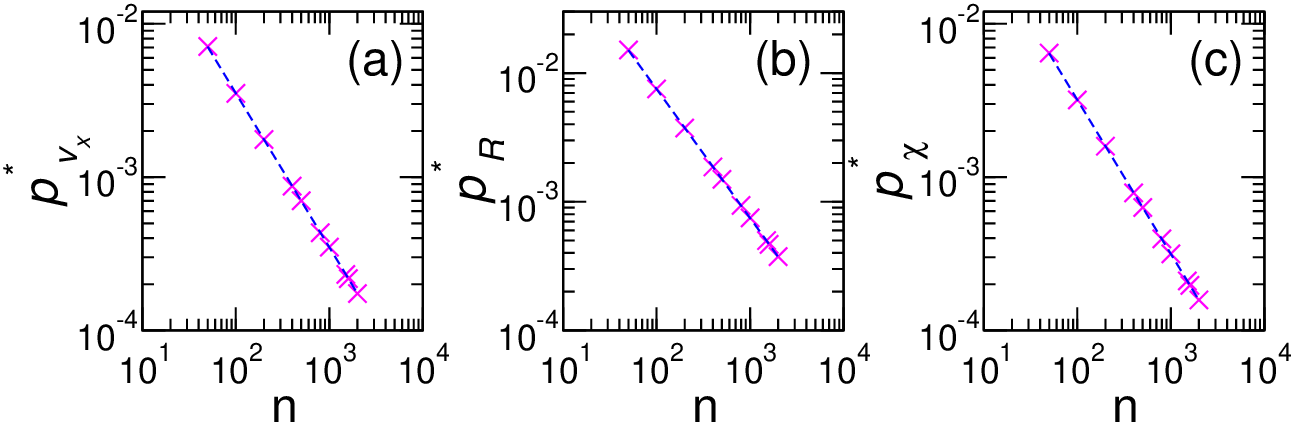}
\caption{$p^*$ for (a) $\bra V_{x}(D) \ket$, (b) $\bra R(D) \ket$ and (c) $\bra \chi(D) \ket$ as a function of the graph size $n$ of Erd\H{o}s-R\'{e}nyi digraphs.}
\label{Fig3}
\end{figure}

In Fig.~\ref{Fig3} we present $p^*$ as a function of the graph size $n$ and observe  a linear trend of the data sets $p^{*}$ vs.~$n$ (in log-log scale), suggesting a power--law behavior of the form
\begin{equation}
p^{*} = \mathcal{C}n^{-\beta}.
\label{pa}
\end{equation}
Then, by performing numerical fittings, we determined the parameters $\mathcal{C}$ and $\beta$ which are reported in Table~\ref{tab1}. There, we can clearly see that $\beta \approx 1$ in all three cases.
Hence, we define the scaling parameter $\xi$ as the ratio $p/p^*$,
\begin{equation}
\xi =\frac{p}{p^*} \propto \frac{p}{n^{\beta}} \propto \frac{p}{n^{-1}} =np.
\end{equation}

\begin{table}
\begin{center}\caption{Values of the constants $\mathcal{C}$ and $\beta$ obtained by fittings of Eq.~(\ref{pa}) to the data in Fig.~\ref{Fig3}.}
\begin{tabular}{| c | c  c c |}\hline
& $\bra V_x(D) \ket$ & $\bra R(D) \ket$ & $\bra \chi(D) \ket$ \\ \hline
$\mathcal{C}$ & $0.3612$ & $0.7608$ & $0.3271$ \\ \hline
$\beta$       & $1.0051$  & $1.0026$  & $1.005$ \\  \hline
\end{tabular}
\label{tab1}
\end{center}
\end{table}

Previous studies on undirected ER graphs have demonstrated that topological measures can be scaled with the average degree $\bra k \ket$~\cite{MAMRP15, MMRS20, MMRS21}. Here, the average degree of ER digraphs is given by
\begin{equation}
\bra k \ket = 2 (n-1)p.
\end{equation}

\begin{figure}
\centering
\includegraphics[scale=0.4]{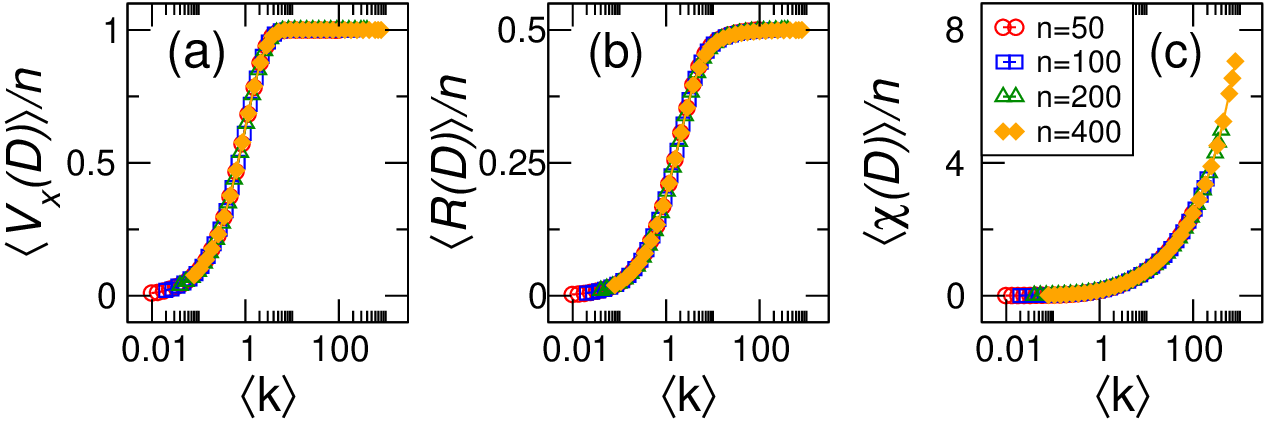}
\caption{(a) $\bra V_{x}(D) \ket$, (b) $\bra R(D) \ket$ and (c) $\bra \chi(D) \ket$ normalized to $n$ as a function of the average degree $\bra k \ket$ of Erd\H{o}s-R\'{e}nyi digraphs of different sizes $n \in [50,400]$. Same data sets of Fig.~\ref{Fig1}. }
\label{Fig4}
\end{figure}

In addition, we can observe that both $\bra k \ket$ and $\xi$ depend on $n$ and $p$ in the same functional form. Therefore, we can express $\xi$ as a function of $\bra k \ket$ and vice versa. Also, it is important to recall that the scaling parameter is not unique; a function of it can also serve as scaling parameter. These observations allow us to propose the average degree $\bra k \ket$ as the scaling parameter for the topological properties of ER digraphs. Then, in Fig.~\ref{Fig4} we present again the curves of Fig.~\ref{Fig2} but now plotted as a function of $\bra k \ket$. As observed, the average degree indeed serves as the scaling parameter of these topological quantities.

\begin{figure}
\centering
\includegraphics[scale=0.4]{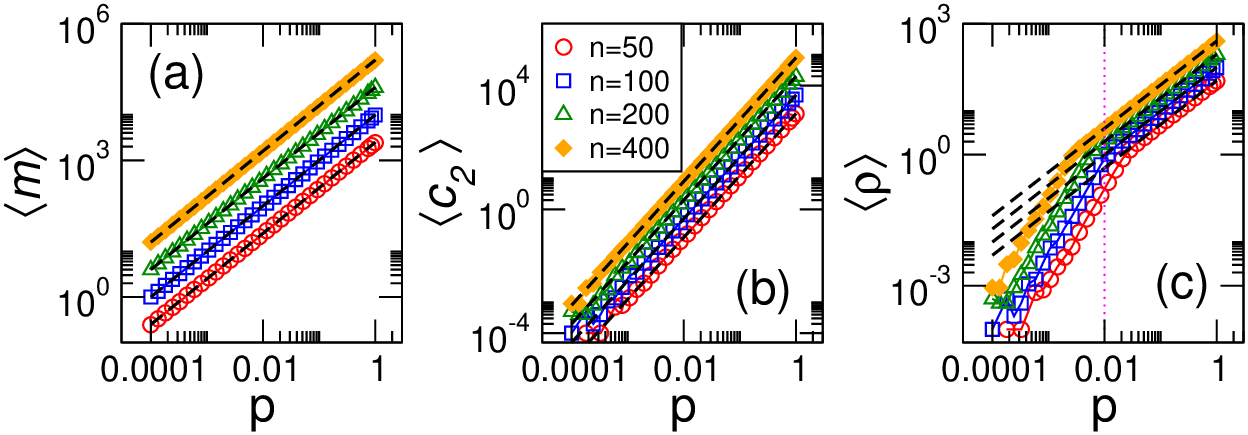}
\caption{(a) Average number of edges $\bra m \ket$, (b) average number of closed walks of length 2 $\bra c_{2} \ket$ and (c) average spectral radius $\bra \rho \ket$ as a function of the connection probability $p$ of Erd\H{o}s-R\'{e}nyi digraphs of different sizes $n \in [50,400]$. Dashed lines in panels (a-c) correspond to $\bra m \ket = n^2p$, $\bra c_2 \ket = n^2p^2/2$, and $\bra \rho \ket = np$, respectively. Each symbol was calculated by averaging over $10^6/n$ random digraphs.}
\label{Fig5}
\end{figure}

Other important quantities in the study of digraphs are the number of arcs $m$, the number of closed walks of length two $c_{2}$, and the spectral radius $\rho$ (the maximum of the absolute values of the adjacency matrix eigenvalues). Our next goal is to compute these quantities and examine whether they can also be scaled with the average degree. To achieve this, we construct ensembles of ER digraphs characterized by different combinations of parameters and compute the average of the quantities above. In Fig.~\ref{Fig5} we plot $\bra m \ket$, $\bra c_2 \ket$ and $\bra \rho \ket$ as a function of the connection probability $p$. Remarkably, these quantities exhibit a  behavior similar to that reported for the previously studied topological indices: Curves representing the same quantity show a similar pattern but they are shifted along the $p$-axis for different graph sizes. This observation strongly suggests that these quantities may also be scaled with the average degree. Furthermore, Fig.~\ref{Fig5} reveals noteworthy characteristics. Specifically, in the case of $\bra m \ket$ and $\bra c_2 \ket$, we observe a linear trend with $p$ on a log-log scale. Numerical calculations indicate that $\bra m \ket$ follows the relationship $\bra m \ket \approx n^2p$. Similarly, for $\bra c_2 \ket$ we find that $\bra c_2 \ket \approx n^2p^2/2$. Additionally, for $p>0.01$, we find that $\bra \rho \ket \approx np$. These approximations, where the average degree can be easily identified (i.e.~$np \approx \bra k \ket/2$), are indicated in each panel of Fig.~\ref{Fig5} with dashed lines.

\begin{figure}
\centering
\includegraphics[scale=0.4]{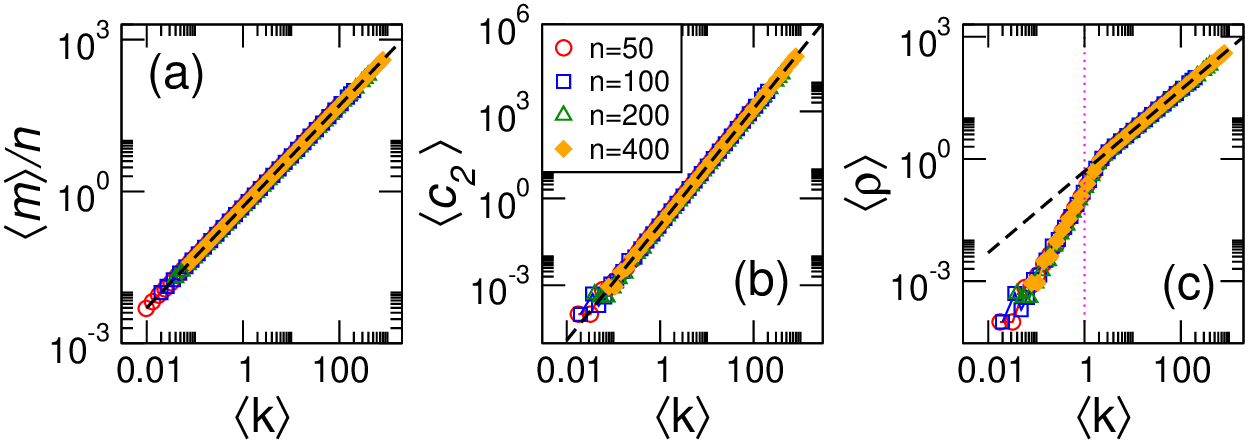}
\caption{(a) $\bra m \ket /n$, (b) $\bra c_{2} \ket$ and (c) $\bra \rho \ket$ as a function of the average degree $\bra k \ket$ of Erd\H{o}s-R\'{e}nyi digraphs of different sizes $n \in [50,400]$. Dashed lines in panels (a-c) correspond to $\bra m \ket /n= \bra k \ket /2$ , $\bra c_{2} \ket = \bra k \ket^2 /8 = m^2/2n^2$ and $\bra \rho \ket = \bra k \ket /2$ , respectively. Same data sets of Fig.~\ref{Fig5}.}
\label{Fig6}
\end{figure}

Then, in Fig.~\ref{Fig6} we verify that the average degree indeed scales $\bra m \ket /n$, $\bra c_{2} \ket$ and $\bra \rho \ket$. Therefore we can finally write $\bra m \ket /n \approx \bra k \ket /2$, $\bra c_{2} \ket \approx \bra k \ket^2 /8 \approx m^2/2n^2$ and $\bra \rho \ket \approx \bra k \ket /2$ for $k>1$; see the dashed lines in the corresponding panels of Fig.~\ref{Fig6}.
These findings provide highly relevant information about the relationships and scaling behavior of topological quantities, such as $m$ and $c_2$, and the spectral measure $\rho$, in relation to the graph parameters $p$, $n$ and $\bra k \ket$. Moreover, the observation that the spectral radius scales with the average degree suggests that other spectral magnitudes could also scale with $\bra k \ket$, which is precisely what is addressed in the next section.

\section{Energy of Erd\H{o}s-R\'{e}nyi digraphs}
\label{energy}

\subsection{Short review of digraph energies}

As mentioned in the Introduction, the definition of energy proposed by Gutman (see Eq.~(\ref{Eq1})) cannot be directly applied to digraphs since, in this case, the eigenvalues can be complex. However, by examining Eq.~(\ref{Eq1}), a straightforward generalization can be done by replacing the absolute value of the real eigenvalues with the module of the complex eigenvalues of the adjacency matrix of a digraph, here denoted as $Z_k$. In fact, this definition of energy,
\begin{equation}
\mathcal{S}(D)= \sum_{k=1}^{n} |Z_{k}|,
\label{eeq1}
\end{equation}
has been reported in Ref.~\cite{MBG10}.
Interestingly, this definition is not the most widely studied. Instead of opting for this direct generalization, other definitions have received more attention.

In Ref.~\cite{PR08}, Pe\~{n}a and Rada, motivated by Coulson's formula, generalized the concept of energy of a digraph as
\begin{equation}
e(D) = \sum_{k=1}^{n} \abs{\operatorname{Re}(Z_{k}) }.
\end{equation}
This digraph energy has been extensively studied; see for example Refs.~\cite{R09,B10,R10,R12, PBGR13}. Also, this definition has been extended to other graph energies~\cite{G12, LSG12}. Bounds have also been established for $e(D)$. For example, in~\cite{R09} Rada generalized McClelland's inequality for directed graphs with $n$ vertices, $m$ edges and $c_2$ closed walks of length $2$ as
\begin{equation}\label{emax}
e(D) \leq \sqrt{\frac{1}{2}n\left(m+c_{2}\right)}.
\end{equation}
Also, a lower bound for $e(D)$ was established~\cite{B10,R10}:
\begin{equation}\label{emin}
e(D) \geq \sqrt{2c_{2}}.
\end{equation}
It is important to note that the definition of Pe\~{n}a and Rada, although it satisfies the Coulson integral, does not consider the imaginary part of the eigenvalues. 

Another definition of energy of digraphs that does consider the imaginary part of the eigenvalues was proposed by Khan, Farooq and Rada in Ref.~\cite{KFR17}. This is called the iota energy and is defined as
\begin{equation}
E_{\iota}(D)= \sum_{k=1}^{n} \abs{\operatorname{Im}(Z_{k})}.
\end{equation}
This energy can be defined from the Coulson integral formula using the characteristic polynomial of the complex adjacency matrix $\mathbf{A}_c$. Which is defined as
\begin{equation}
\mathbf{A}_{c_{u v}}=\left\{\begin{array}{ll}
-\imath  & \text { if } u \rightarrow v, \\
0 & \text { otherwise. }
\end{array}\right.
\end{equation}
The iota energy has been extensively studied in digraphs with specific characteristics such as bicyclic, tricyclic, and signed digraphs~\cite{FKA17, YW18, FKC19, FCK19, FK18}.

More recently, Khan proposed another energy definition that incorporates both real and imaginary parts of the adjacency matrix eigenvalues. This energy is called the p-energy and is defined as~\cite{K21, KKA21}
\begin{equation}
E_{p}(D)= \sum_{k=1}^{n} \abs{\operatorname{Re}(Z_{k}) \operatorname{Im}(Z_{k})}.
\end{equation}
To represent this energy in an integral way with Coulson's formula it is necessary to use the characteristic polynomial of the squared adjacency matrix $\mathbf{A}^2$ instead of the characteristic polynomial of $\mathbf{A}$.

Moreover, Nikiforov~\cite{N07} proposed the concept of energy of a matrix using the corresponding singular values. The singular values of a matrix are a set of non-negative elements that are calculated from a matrix $\mathbf{A} \in \mathcal{R}^{m\times n}$. They are defined as the square root of the eigenvalues of the $ \mathbf{A}^{T} \mathbf{A} \in \mathcal{R}^{	n\times n}$ matrix.  Given the singular values of a matrix $\mathbf{A}$, $\sigma_k$, the energy of the matrix is defined as
\begin{equation}
\mathcal{N}(\mathbf{A})= \sum_{k=1}^{n}\sigma_{k}.
\end{equation}
This concept has been widely studied for different types of matrices, such as non-square matrices~\cite{KT08} and for digraphs~\cite{AR16, GMR21}. Its importance lies in the fact that it can be computed for any matrix and, in the case of a square symmetric matrix, it reproduces Eq.~(\ref{Eq1}). For the Nikiforov energy, some bounds have been reported in terms of the properties of the matrix. Particularly, an upper bound for $\mathcal{N}(D)$ has been reported as ~\cite{B10,KT08}
\begin{equation}\label{c3}
\mathcal{N}(D) \leq  \frac{m}{n}+\sqrt{\left(n-1\right)\left(m-\frac{m^2}{n^2}\right)},
\end{equation}
while Agudelo and Rada proposed the lower bound~\cite{AR16}
\begin{equation}\label{c4}
\mathcal{N}(D) \geq \sqrt{m}.
\end{equation} 

Another definition of energy reported in the literature is the Hermitian energy~\cite{LL15, GM17, BP22}. In contrast to the previous definitions, this energy is not directly linked to the adjacency matrix of the digraph. To calculate this energy, it is necessary to construct the Hermitian adjacency matrix, denoted as $\bf{H}$, which is defined as follows:
\begin{equation}
H_{u v}=\left\{\begin{array}{ll}
1 & \text { if } u \leftrightarrow v, \\
-\imath  & \text { if } u \rightarrow v, \\
\imath  & \text { if } v \rightarrow u, \\
0 & \text { otherwise. }
\end{array}\right.
\end{equation}
Since $\bf{H}$ is a Hermitian matrix by construction, its eigenvalues are real. Then, the Hermitian energy, denoted as $E_H(G)$, can be computed using Eq.~(\ref{Eq1}) with the eigenvalues of $\bf{H}$. Bounds have also been established for $E_H(G)$. Considering $q$ as the determinant of $\bf{H}$ and $\Delta$ the maximum degree of the graph, the following bounds were derived~\cite{LL15}:
\begin{equation}
\sqrt{2m+n(n-1)q^{2/n}}\leq E_{H}(G) \leq n \sqrt{\Delta}.
\label{eh1}
\end{equation} 
Additionally, when considering solely the number of arcs, an alternative bound for the Hermitian energy is expressed as~\cite{LL15}
\begin{equation}
2\sqrt{m} \leq E_{H}(G) \leq 2m.
\label{eh2}
\end{equation} 
Particularly for ER digraphs, considering the relationships of $m$, $n$, and $p$ found in the previous section, this bound can be expressed as
\begin{equation}
2n\sqrt{p} \leq E_{H}(G) \leq 2n^2p.
\label{eh3}
\end{equation}

Although all these energy definitions have been extensively studied to determine minimum and maximum bounds and have also been computed for specific graphs, no numerical study has yet been performed to compare them. To fill this gap, we have undertaken the task of numerically and statistically evaluating these energies for ensembles of ER random digraphs.

\subsection{Energies of ER digraphs}

Here we compute the ${\cal S}(D)$ energy, the Pe\~{n}a-Rada energy $e(D)$, the iota energy $E_{\iota}(D)$, the p-energy $E_{p}(D)$, the Nikiforov energy $\mathcal{N}(D)$, and the Hermitian energy $E_{H}(D)$ for ensembles of ER digraphs characterized by the parameter pair $(n,p)$.

\begin{figure}
\centering
\includegraphics[scale=0.42]{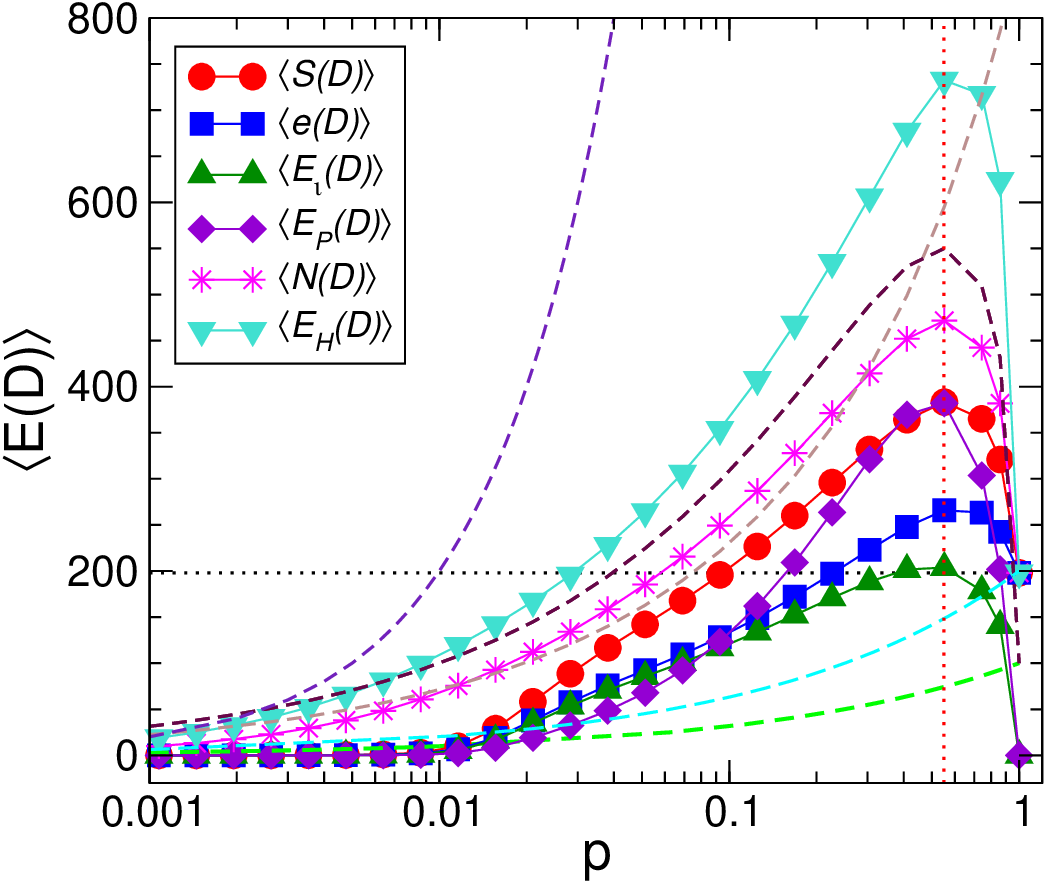}
\caption{Average energies as a function of the connection probability  $p$ of Erd\H{o}s-R\'{e}nyi digraphs of size $n=100$. The dashed lines indicate the limits defined by equations (\ref{emax}) (brown color), (\ref{c3}) (maroon color), (\ref{c4}) (green color) and the upper and lower bounds given by Eq.~(\ref{eh3}) (violet and cyan, respectively). The black dotted line indicates the hyperenergetic limit. The vertical red-dotted line is $p=\frac{1}{2}(1+\frac{1}{\sqrt{n}})$ with $n=100$. Each symbol was calculated by averaging over $10^6/n$ random digraphs.}
\label{Fig7}
\end{figure}

Then, in Fig.~\ref{Fig7} we plot the average energies of ER digraphs of size $n=100$ as a function of the connection probability $p$. In Fig.~\ref{Fig7} we also indicate the bounds given by Eqs.~(\ref{emax}), (\ref{c3}), (\ref{c4}) and (\ref{eh3}), and the hyperenergetic limit $2n-2$.

From Fig.~\ref{Fig7}, we can see that all energies exhibit a similar pattern as a function of $p$: As $p$ increases, the energy  increases until reaching a maximum value at $p$ close to 1; then, it decreases displaying a bell-like shape that is better observed in a semi-logarithmic scale. However, the maximum values for different energies are different. We numerically computed the maximum values reached by the different energies for ensembles of digraphs of different sizes (not shown here). We found that all energies reach their maximum at $p \approx 0.5$. Moreover, we can see in Fig.~\ref{Fig7} that the curve corresponding to Eq.~(\ref{c3}) also reaches its maximum at $p \approx 0.5$, in agrrement with all the numerically computed energies. Then, in order to get an estimation of the value of $p$ producing the energy maxima we rewrite Eq.~(\ref{c3}) in terms of $n$ and $p$ using $m\approx n^2p$, then we see that 
\begin{eqnarray}
\mathcal{N}(D) \leq np + \sqrt{(n-1)(n^2p-n^2p^2)}\nonumber\\
	= \ np + n\sqrt{(n-1)p(1-p)}.
\label{c5}
\end{eqnarray}
So we find that the maximum of Eq.~(\ref{c5}) occurs at $p=\frac{1}{2}(1+\frac{1}{\sqrt{n}})$, which is consistent with the numerical observation.

\begin{figure}
\centering
\includegraphics[scale=0.405]{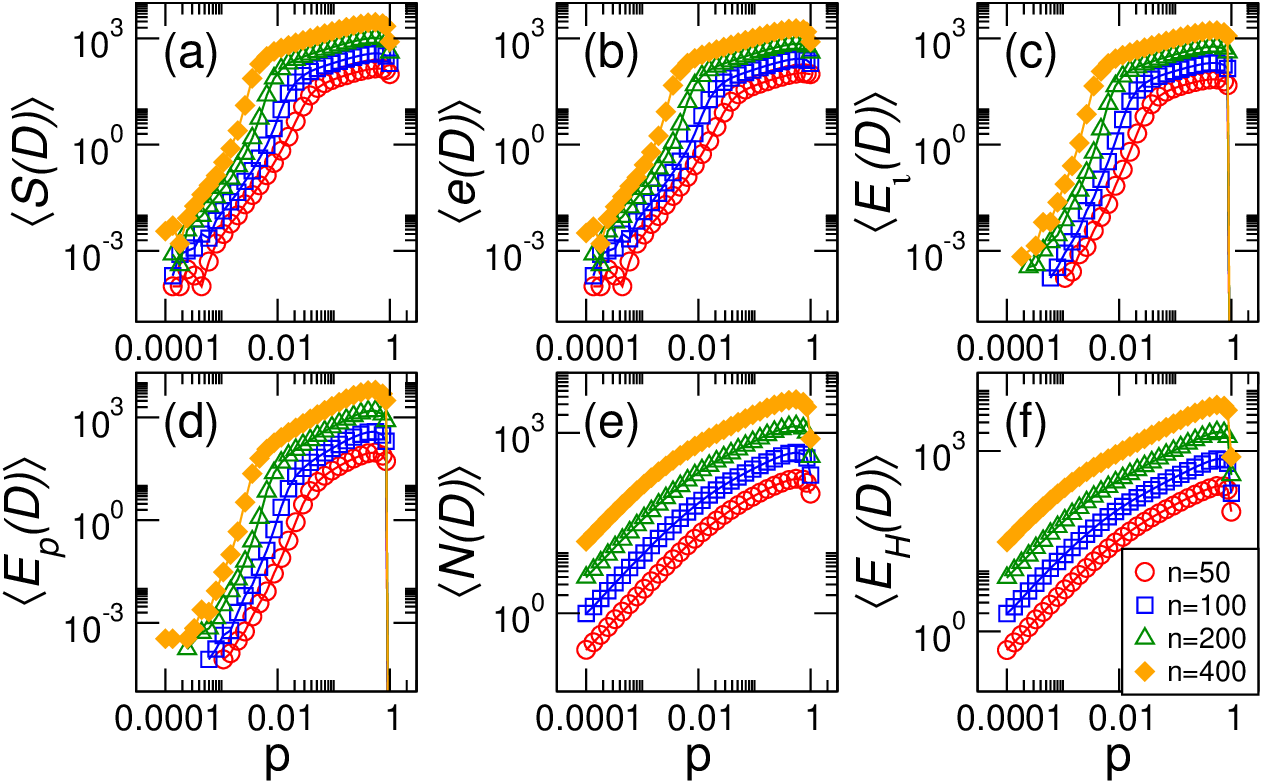}
\caption{Average energy $\bra {\cal S}(D) \ket$, (b) average Pe\~{n}a-Rada energy $\bra e(D) \ket$, (c) average iota energy $\bra E_{\iota}(D)\ket $, (d) average p-energy $\bra E_{p}(D)\ket $, (e) average Nikiforov energy $\bra \mathcal{N}(D)\ket $ and (f) average hermitian energy $\bra E_{H}(D) \ket $ as a function of the connection probability $p$ of Erd\H{o}s-R\'{e}nyi digraphs of sizes $n \in [50,400]$. Each symbol was calculated by averaging over $10^6/n$ random digraphs.}
\label{Fig8}
\end{figure}

\begin{figure}
\centering
\includegraphics[scale=0.4]{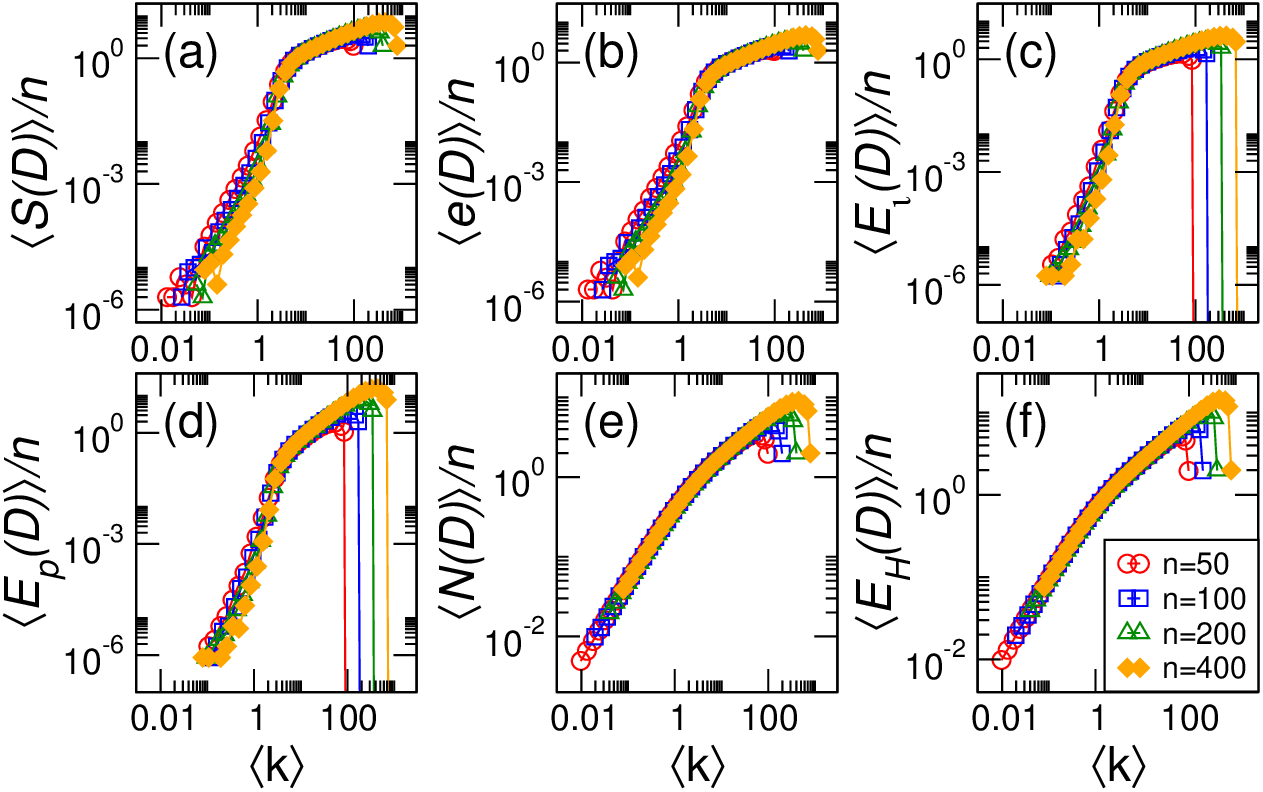}
\caption{(a) $\bra {\cal S}(D) \ket$, (b) $\bra e(D) \ket$, (c) $\bra E_{\iota}(D)\ket $, (d) $\bra E_{p}(D)\ket $, (e) $\bra \mathcal{N}(D)\ket $ and (f) $\bra E_{H}(D) \ket $ normalized to $n$ as a function of the average degree $\bra k \ket$ of Erd\H{o}s-R\'{e}nyi digraphs of sizes $n \in [50,400]$. Same data sets of Fig.~\ref{Fig8}.}
\label{Fig9}
\end{figure}

We recall that in Fig.~\ref{Fig7} we used ER digraphs of size $n=100$ so, to see the effect of the graph size 
on the average energies, in Fig.~\ref{Fig8} we plot them as a function of $p$ for different values of $n$.
From this figure we can observe that the curves for a given energy definition exhibit a similar functional 
dependence of $p$ but they are shifted on both axis for increasing $n$. 
This effect of $n$ on the energies is equivalent to that observed in the previous Section for the topological and
spectral properties of ER digraphs, see Figs.~\ref{Fig1} and~\ref{Fig5}. 
Thus, taking as a reference the scaling analysis of Sec.~\ref{topological}, in Fig.~\ref{Fig9} we plot the
energies normalized to $n$ now as a function of the average degree.
Indeed, we observe that $\bra k \ket$ works well as the scaling parameter of the normalized energies,
mainly above the percolation threshold $\bra k \ket > 1$. Moreover, remarkably, $\bra k \ket$ perfectly 
scales the normalized Nikiforov energy as well as the normalized Hermitian energy over the entire range 
of connection probabilities, see Figs.~\ref{Fig9}(e,f).

\begin{figure}
\centering
\includegraphics[scale=0.4]{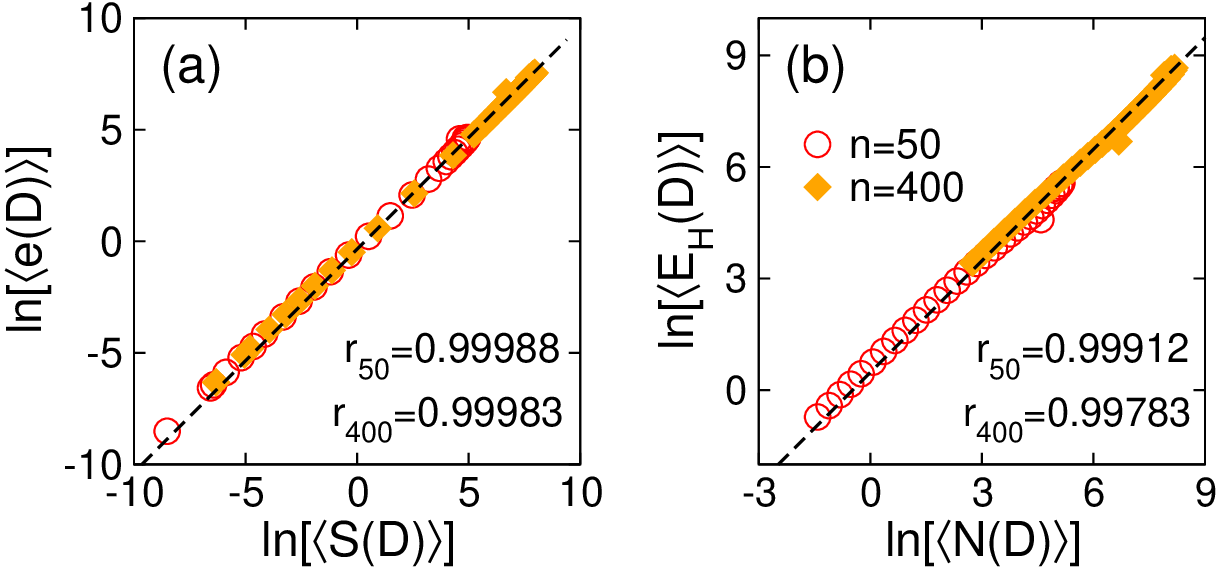}
\caption{Scatter plots of (a) $\bra {\cal S}(D) \ket$ vs.~$\bra e(D) \ket$ and
(b) $\bra \mathcal{N}(D) \ket$ vs.~$\bra E_{H}(D) \ket$. Data corresponds to $n=50$ and 400.
The Pearson correlation coefficients $r$ are reported in the corresponding panels.
The black-dashed lines are fittings of the form $y={\cal C} x$ with (a)  ${\cal C}=1/\sqrt{2}$, and (b) ${\cal C}=\frac{8}{5}$. }
\label{Fig10}
\end{figure}

In addition, in Fig.~\ref{Fig9} we can see that certain energy pairs depend on $\bra k \ket$ in a very similar 
way and, consequently, they should be strongly correlated.
Specifically, we observe strong similarities between ${\cal S}(D)$ and $e(D)$, see Figs.~\ref{Fig9}(a,b); 
 and between $\mathcal{N}(D)$ and 
$E_{H}(D)$, see Figs.~\ref{Fig9}(e,f).
So, in Fig.~\ref{Fig10} we present scatter plots of these pairs of energies and report the corresponding 
Pearson correlation coefficients. 
Moreover, the strong correlations reported in Fig.~\ref{Fig10} allowed us to state the following relations:
\begin{eqnarray}
\label{Svse}
\sqrt{2} \ e(D) & \approx & {\cal S}(D) , \\
\label{NvsH}
E_{H}(D) & \approx & \frac{8}{5} \ \mathcal{N}(D) ,
\end{eqnarray}
see the black-dashed lines Fig.~\ref{Fig10}.

\section{Conclusions and discussion}
\label{conclusions}

This study aims to contribute to the understanding of topological and spectral properties of random
digraphs. Specifically, we studied some topological and spectral properties of Erd\H{o}s-R\'{e}nyi (ER) 
digraphs $D(n,p)$.

Initially, we focused on the statistical analysis of topological properties by computing the average number 
of non-isolated vertices $\bra V_{x}(D) \ket$, the average Randi\'c index $\bra R(D) \ket$ and the average 
sum-connectivity index $\bra \chi(D) \ket$. 
By means of a scaling analysis, we found the total average degree $\bra k \ket$ works well as scaling 
parameter of  $\bra V_{x}(D) \ket$, $\bra R(D) \ket$ and $\bra \chi(D) \ket$ but also for the average number
of arcs $\bra m(D) \ket$, the average spectral radius $\bra \rho(D) \ket$ and the average closed walks of 
length 2 $\bra c_2(D) \ket$.
Moreover, we were able to infer the following relations:
$\bra m(D) \ket/n \approx \bra k \ket /2$, 
$\bra c_2(D) \ket \approx \bra k \ket^2 /8$, and 
$\bra \rho(D) \ket \approx \bra k \ket /2$ for $\bra k \ket>1$.

Concerning spectral properties, we computed six different graph energies for ensembles of ER digraphs 
$D(n,p)$:
the ${\cal S}(D)$ energy, the Pe\~{n}a-Rada energy $e(D)$, the iota energy $E_{\iota}(D)$, the p-energy 
$E_{p}(D)$, the Nikiforov energy $\mathcal{N}(D)$, and the Hermitian energy $E_{H}(D)$.
First, we showed that $\bra k \ket$ scales well all normalized averaged energies,
mainly above the percolation threshold $\bra k \ket > 1$. Moreover, remarkably, $\bra k \ket$ perfectly 
scales $\bra \mathcal{N}(D) \ket/n$ and $\bra E_{H}(D) \ket/n$ over the entire range 
of connection probabilities.
Then, we reformulated a set of bounds previously reported in the literature for these energies as a function 
$(n,p)$.
So, by identifying strong correlations between ${\cal S}(D)$ and $e(D)$, and between $\mathcal{N}(D)$ and $E_{H}(D)$
we phenomenologically stated linear relations between energies, see Eqs.~(\ref{Svse}-\ref{NvsH}).

\begin{figure}
\centering
\includegraphics[scale=0.36]{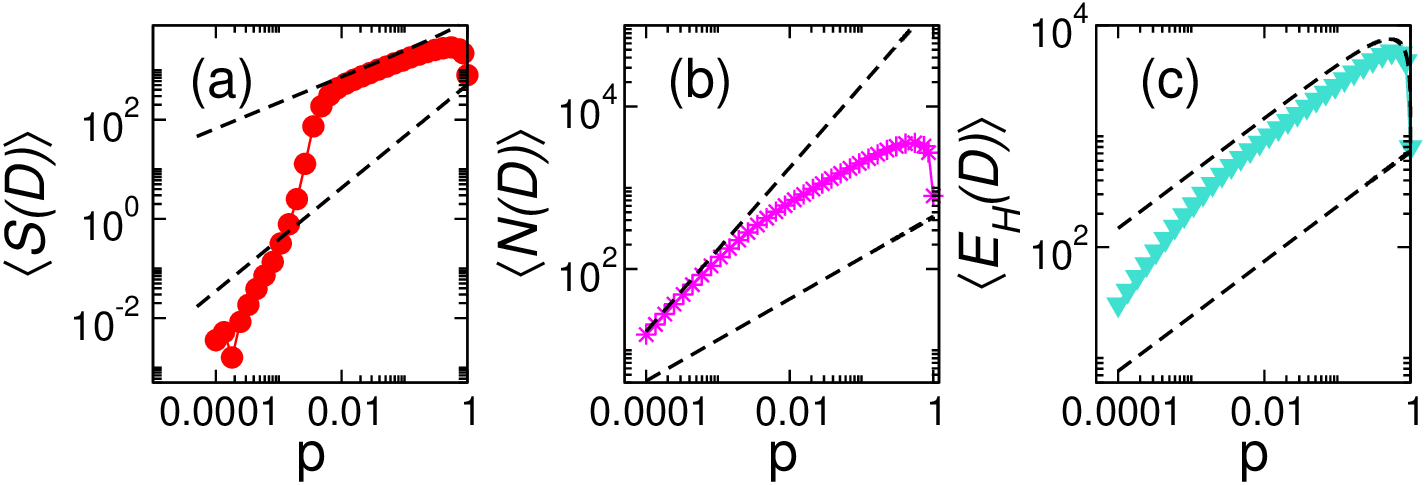}
\caption{(a) $\bra {\cal S}(D) \ket$, (b) $\bra \mathcal{N}(D)\ket $ and (c) $\bra E_{H}(D) \ket$ as a function of the connection probability $p$ of Erd\H{o}s-R\'{e}nyi digraphs of size $n=400$. The dashed lines in panels (a), (b) and (c) indicate the limits given by Eqs.~(\ref{eq31}), (\ref{eq32}) and~(\ref{eq33}), respectively.}
\label{Fig11}
\end{figure}

It is important to stress that Eqs.~(\ref{Svse}-\ref{NvsH}) can be used to extend previously known bounds.
That is, from Eqs.~(\ref{emax}-\ref{emin}) and~(\ref{Svse}) we get
$$
2\sqrt{c_{2}} \leq S(D) \leq \sqrt{n\left(m+c_{2}\right)},
$$
by combining Eqs.~(\ref{c3}-\ref{c4}) and~(\ref{NvsH}) we can write 
$$
\frac{8}{5} \sqrt{m}\leq E_{H}(D) \leq \frac{8}{5} \left[\frac{m}{n} + \sqrt{\left(n-1\right)\left(m-\frac{m^2}{n^2}\right)} \right]
$$
while from Eq.~(\ref{eh2}) and~(\ref{NvsH}) we obtain
$$
\frac{5}{4} \sqrt{m}\leq \mathcal{N}(D)\leq \frac{5}{4}m.
$$
Which in the particular case of ER digraphs, read as:
\begin{equation}\label{eq31}
\sqrt{2}np \leq S(D) \leq n \sqrt{np\left(1+\frac{p}{2}\right)},
\end{equation}
\begin{equation}\label{eq32}
\frac{8}{5} n\sqrt{p}\leq E_{H}(D) \leq \frac{8}{5}n \left[p +\sqrt{\left(n-1\right) p \left(1-p \right)} \right]
\end{equation}
and
\begin{equation}\label{eq33}
\frac{5}{4} n\sqrt{p}\leq \mathcal{N}(D)\leq \frac{5}{4}n^2p,
\end{equation}
respectively.
Finally, in Fig.~\ref{Fig11}, we validate Eqs.~(\ref{eq31}-\ref{eq33}) on ER digraphs of size $n=400$.
We just note that the lower bound in Eq.~(\ref{eq31}) fails to bound $\bra {\cal S}(D) \ket$, see
Fig.~\ref{Fig11}(a); however this also happens in the original Eq.~(\ref{emin}).

\begin{acknowledgments}
J.A.M.-B. thanks support from CONACyT (Grant No. 286633), CONACyT-Fronteras (Grant No. 425854), and
VIEP-BUAP (Grant No. 100405811-VIEP2023), Mexico.
C. T. M. M. thanks support from CONAHCYT (CVU No. 784756).
\end{acknowledgments}

\end{document}